\documentclass[twocolumn]{pasj01}

\usepackage{lineno}

\Received{}
\Accepted{}
 
\usepackage[]{graphicx}
\usepackage{subfigure}

\begin{document} 

\title{ 
Very Long Baseline Interferometry imaging 
of H$_2$O maser emission 
in the nearby radio galaxy NGC~4261 
}

\author{Satoko \textsc{Sawada-Satoh}\altaffilmark{1}%
}
\altaffiltext{1}{
Graduate School of Science, Osaka Metropolitan University, 
1-1 Gakuen-cho, Naka-ku, Sakai, Osaka 599-8531, Japan 
}
\email{sss@omu.ac.jp}

\author{Nozomu \textsc{Kawakatu},\altaffilmark{2}}
\altaffiltext{2}{
National Institute of Technology, Kure College, 
2-2-11, Agaminami, Kure, Hiroshima, 737-8506, Japan
}

\author{Kotaro \textsc{Niinuma}\altaffilmark{3,4}}
\altaffiltext{3}{
Graduate School of Sciences and Technology for Innovation, 
Yamaguchi University, 
1677-1 Yoshida, Yamaguchi-city, Yamaguchi 753-8512, Japan
}
\altaffiltext{4}{
The Research Institute for Time Studies, 
Yamaguchi University, 
1677-1 Yoshida, Yamaguchi 753-8511, Yamaguchi Japan
}

\author{Seiji \textsc{Kameno}\altaffilmark{5,6}}
\altaffiltext{5}{
Joint ALMA Observatory, 
Alonso de Córdova 3107, Vitacura, Santiago, 763-0355, Chile
}
\altaffiltext{6}{
National Astronomical Observatoryh of Japan, 
2-21-1 Osawa, Mitaka, Tokyo 181-8588, Japan
}


\maketitle

\begin{abstract}
We report dual-frequency 
very long baseline interferometry (VLBI) observations 
at 22 and 43~GHz 
toward the nucleus of 
a nearby radio galaxy NGC~4261.  
In particular, we present a VLBI image of the 22~GHz H$_2$O maser line   
and its location in the circumnuclear region of NGC~4261. 
H$_2$O maser emission is marginally detected 
above the three times the rms level 
within a velocity range of approximately 2250--2450 km~s$^{-1}$, 
slightly red-shifted 
with respect to the systemic velocity. 
H$_2$O maser emission is located 
approximately 1 milliarcsecond (mas) east of 
the brightest continuum component at 22~GHz,
where the continuum spectrum is optically thick,  
that is at the free--free absorbed receding jet 
by ionized gas. 
A positional coincidence 
between H$_2$O maser emission and an ionized gas disk 
implies that the 
H$_2$O maser emission arises 
from the near side of the disk, 
amplifying continuum emission 
from the background receding jet. 
If the disk axis is oriented 64$^{\circ}$ 
relative to the line of sight, 
the H$_2$O maser emission is expected to be 
at a mean radius of 0.3~pc in the disk. 
The broad line width of the H$_2$O maser emission 
can be attributed to complex kinematics 
in the immediate vicinity of 
the supermassive black hole (SMBH), 
including ongoing gas infall onto the SMBH, 
turbulence, and outflow. 
This is analogous to the multi-phase circumnuclear torus model 
in the nearest radio-loud H$_2$O megamaser source NGC~1052. 
An alternative explanation for H$_2$O maser association 
is the shock region between the jet and the ambient molecular clouds.  
However, this explanation 
fails to describe the explicit association of
H$_2$O maser emission only 
with the free--free absorbed receding jet. 

\end{abstract}

\section{Introduction}

Luminous extragalactic H$_2$O maser (megamaser) emission
in the 6$_{16}$--5$_{23}$ rotational transition 
at 22.235 GHz 
is known to 
originate within a few parsec (pc)
from the supermassive black hole (SMBH) of the 
active galactic nucleus (AGN). 
Notably, studies on megamasers have proven to 
be powerful tools   
in investigating circumnuclear gas distribution and 
kinematics on pc scales. 
So far, 
megamasers have been interpreted to be associated with 
several AGN phenomena, such as 
edge-on accretion disks  
(e.g. \cite{miyoshi95,herrnstein05,reid09,kuo11}), 
interactions between radio jets and 
ambient molecular clouds
(e.g. \cite{claussen98,peck03}), 
and nuclear outflows 
(e.g. \cite{greenhill03, kondratko05}).

To date, H$_2$O maser emission at 22~GHz 
has been found in nearly 200 galaxies including 150 AGNs
\footnote{
https://safe.nrao.edu/wiki/bin/view/Main/PublicWaterMaserList}
(e.g. \cite{braatz18, pesce23}), 
which are predominantly classified as 
low-ionization nuclear emission regions (LINERs) 
or Seyfert 2s (e.g., \cite{braatz97})
except for a Seyfert 1 
\citep{hagiwara03}. 
The radio continuum emissions from these AGNs 
are generally weak
\citep{braatz97}, 
and they are mostly associated with 
radio-quiet AGNs. 
Moreover, 
H$_2$O megamasers have rarely been detected 
in radio-loud objects such as 
NGC~1052 \citep{braatz94, claussen98},  
TXS 2226-184 \citep{koekemoer95,surcis20}, 
3C 403 \citep{tarchi03, tarchi07}, 
Mrk 348 \citep{peck03}
and NGC~4261 \citep{wagner13}. 
High-resolution observations 
of the nearest ($z =$ 0.005) 
radio-loud H$_2$O megamaser source, NGC~1052,  
have presented a scenario in which 
H$_2$O maser emission has been observed to arise  
from molecular clouds
in a circumnuclear torus \citep{sss08}.
This scenario explains that 
the H$_2$O molecules are excited 
at a temperature of 400 K 
in an X-ray dissociation region 
(XDR; \cite{maloney02}) 
inside the torus or disk 
and amplify the the continuum seed emission 
from bright nuclear jets in the background. 
However, 
whether the H$_2$O maser emission associated with 
a pc-scale torus in NGC~1052 is common or unique 
among the radio-loud H$_2$O megamaser sources 
remains unclear, 
because high-resolution observations 
of such sources 
have often been highly challenging 
owing to the distances to these sources and 
their weak emissions. 
Therefore, 
there is a pressing need to determine 
the location and kinematics of the H$_2$O maser gas 
in the second nearest ($z =$ 0.0074) 
radio-loud H$_2$O megamaser source NGC~4261.

NGC~4261 (3C 270) is a nearby 
Fanaroff–Riley I radio galaxy 
with a 
symmetric two-sided radio jet 
raning from kpc to pc scales 
along the east-west direction 
\citep{birkinshaw85}. 
It hosts a type 2 LINER AGN 
with a low X-ray luminosity of 
$L_\mathrm{2-10 keV}$ = $10^{41.1}$ erg s$^{-1}$ 
and a high X-ray absorbing column density of 
$N_\mathrm{H}$ = 1.64$\times$10$^{23}$ cm$^{-2}$ 
\citep{gonzalez09}. 
The western and eastern jet  
approach and recede from the observers, 
respectively 
(e.g., \cite{vanlangevelde00, haga15}).
Multi-epoch very long baseline interferometry (VLBI) 
observations by \citet{piner01} 
have measured 
the jet orientation angle ($\theta_\mathrm{jo}$) of ($63\pm3$)$^{\circ}$
with respect to the line of sight,  
and an intrinsic speed of 
($0.46\pm0.02$)$c$ for the radio jet. 
A circumnuclear disk (CND) of gas and dust 
with a radius of a few 100 pc
is imaged using the Hubble Space Telescope
\citep{jaffe93, jaffe96, ferrarese96}, 
and it is also associated with molecular lines  
as revealed 
by recent high resolution interferometric images 
captured at millimeter and submillimeter wave lengths 
\citep{boizelle21, sss22}. 

Past multi-frequency VLBI observations 
have revealed 
free--free absorption (FFA)
of synchrotron emission 
on the base of two-sided jet 
caused by dense ionized gas condensation 
\citep{jones97, jones00, jones01, haga15}. 
In addition to the ionized gas, 
neutral atomic hydrogen (H~$\textsc{i}$)
absorption has been detected 
at the systemic velocity ($V_\mathrm{sys}$) 
of the galaxy 
using the Very Large Array 
\citep{jaffe94},
and it has been confirmed 
at the front of the receding jet 
at 18 mas ($\sim$ 2.5 pc) shifted 
from the core component at 1.4 GHz 
via the European VLBI Network (EVN) 
observations 
\citep{vanlangevelde00}. 
These absorbing gases can be 
in the form of an obscuring disk 
surrounding the SMBH  
at the inner pc-scale radii of the CND. 
Under the assumption that the H~$\textsc{i}$ gas 
is located in the inner region of the CND, 
the disk opening angle ($\phi_\mathrm{do}$) is estimated 
to be $\sim13^{\circ}$ 
using the 
H~$\textsc{i}$ absorption spectral profile 
\citep{vanlangevelde00}. 
Based on phase-referencing VLBI observations,  
\citet{haga15} have measured radio core positions 
showing observing frequency dependences.  
The authors have determined that 
the location of the SMBH is separated by 
0.082 and 0.202 mas 
($=$ 0.12$+$0.082 mas) east 
from the core positions at 43~GHz and 22~GHz, 
respectively. 

\citet{wagner13} first 
detected the 22~GHz H$_2$O maser emission 
from NGC~4261 
using the 100-m Robert C. Byrd Green Bank Telescope (GBT). 
The H$_2$O maser line  
exhibits a broad line profile 
without spike-like narrow-line components. 
The broad line profile is fitted to a single Gaussian 
distribution 
with a full width half maximum (FWHM) 
of 154 km~s$^{-1}$.  
The peak emission is detected at a radial velocity 
of 
approximately at 2300 km~s$^{-1}$, 
which is slightly red-shifted 
relative to the $V_\mathrm{sys}$. 
The total equivalent isotropic luminosity 
is estimated to be 50 $L_{\odot}$, 
significantly more luminous than that observed 
from a typical star-forming region in our Galaxy 
(e.g., \cite{anglada96}). 
The absence of narrow maser lines 
implies the saturated state of the maser 
with low-gain amplification, 
because the narrowing of the maser line represents 
the enhancements in the line center 
owing to unsaturated exponential amplification 
(e.g., \cite{elitzur82}).

In this paper, we present the first image of 
the H$_2$O maser line from NGC~4261 
obtained based on high-sensitivity VLBI observations. 
We adopted a luminosity distance ($D_\mathrm{L}$) 
of 31.7 Mpc and a 
$V_\mathrm{sys}$ 
of 2212 km~s$^{-1}$ (e.g., \cite{babyk19,cappellari11}). 
Note that 1 mas corresponds 
to 0.15 pc for the galaxy.

\begin{table*}
  \tbl{Summary of observations}{%
  \begin{tabular}{ccccccc}
    \hline
    Date & $\nu$ & Stations & $t_\mathrm{on}$ & Synthesized beam & $I_\mathrm{rms}$ & $I_\mathrm{peak}$   \\ 
    (yyyy-mm-dd) & (GHz) & & (hr) & (mas $\times$ mas) & (mJy beam$^{-1}$) & (mJy beam$^{-1}$)  \\
    (1) & (2) & (3) & (4) & (5) & (6) & (7) \\
    \hline
    2021-12-20 & 22 & KVN, VERA, Takahagi & 5.8 & 1.53 $\times$ 1.26, $-16^{\circ}$ & 0.36 & 258 \\
    2021-12-21 & 43 & KVN, VERA & 6.7 & 0.84 $\times$ 0.62, $+23^{\circ}$ & 0.48 & 123 \\
      \hline
    \end{tabular}}
    \label{tab:obsinfo}
\begin{tabnote}
Note. ---(1) Date of observation.  
(2) Frequency. (3) Participating stations.  
(4) Total on-source time. 
(5) Major axis, minor axis and position angle of the synthesized beam.  
(6) Image rms level.  
(7) Peak intensity.  
\end{tabnote}
\end{table*}

\section{Observations and Data Reduction}
 
Dual-frequency VLBI observations of NGC~4261 
were conducted 
on December 20, 2021, at 22~GHz 
and December 21, 2021, at 43~GHz, 
within the framework of the East Asia VLBI Network 
science program. 
The participating stations included 
all stations of the Korean VLBI Network 
(KVN; \cite{lee14}), VLBI Exploration of Radio Astrometry 
(VERA; \cite{kawaguchi00,honma02}), 
and Takahagi 32-m telescope \citep{yonekura16}, 
producing projected baseline lengths 
ranging from 180 to 2261 km. 
Along with NGC~4261, 
bright continuum 
(OJ~287, 3C~273, PKS~1502+106) 
and Galactic maser sources 
(R~Leo, RT~Vir, RS~Vir) 
were observed for 0.1 h every 2 h. 
These were used for fringe finding and several calibrations. 
Data were recorded in left circular polarization 
at a rate of 1024 Mbps, 
with eight spectral windows (SWs) 
and a 32 MHz bandwidth per SW. 
The SWs were set up to cover 
extragalactic 22~GHz H$_2$O maser lines from NGC~4261 
at a velocity of $\sim$ 2200 km~s$^{-1}$ 
and the Galactic H$_2$O maser lines
at velocities from $-20$ to $30$ km~s$^{-1}$. 
The eight SWs were later glued together  
during the data reduction process. 
The recorded data were correlated 
using the Daejeon hardware correlator 
located at the Korea--Japan Correlation Center 
\citep{lee15b}.

Data reduction,  
including calibration, 
data editing, fringe fitting and imaging,  
was performed 
 using the NRAO  
Astronomical Image Processing System 
(AIPS; \cite{Greisen03}) package. 
The visibility amplitude decrement 
attributed to the digital quantization loss was corrected 
by applying a correction factor of 1.3  
\citep{lee15a}. 
A priori amplitude calibration method was performed 
based on known gain curves and system temperature 
measurements recorded individually 
at each station during the observation period 
for all stations except Takahagi. 
For Takahagi, 
the total-power spectra of 
the Galactic H$_2$O maser sources 
were fitted to a best-calibrated template spectrum 
to derive the amplitude calibration factors 
as functions of time. 
The instrumental delays and 
complex bandpass characteristics were calibrated 
using the bright continuum sources. 
Fringe fitting was then performed  
using the continuum channels of NGC~4261, 
to obtain the delay, rate and phase residuals. 
After eliminating these residuals, 
all spectral channels at 43~GHz 
and the line-free channels at 22~GHz 
were integrated into a single continuum channel 
for each band. 
For spectral data, 
the Doppler-shifted velocities 
in H$_2$O maser emission 
attributed to Earth's motion during the observations 
were corrected. 
The continuum emission was subtracted 
in the $uv$ domain 
by fitting to the line-free channels. 
Images were obtained  
using a uniform weighting scheme, 
to achieve higher spatial resolutions. 

We employed an in-beam phase-referencing technique 
between the H$_2$O maser and the continuum emissions 
by applying the solutions of fringe fitting and self-calibration 
derived from the continuum data to the H$_2$O maser data. 
This technique allows to enhance the sensitivity 
of the observations 
and  
to accurately determine 
the relative distribution of 
the H$_2$O maser emission   
with respect to the nuclear continuum source at 22~GHz. 
We created a line-only image cube of H$_2$O maser emission 
with a velocity resolution of 33.7 km~s$^{-1}$, 
after the continuum subtraction. 

\section{Results}

\subsection{Continuum Emission}

\begin{figure}
 \begin{center}
  \includegraphics[width=80mm]{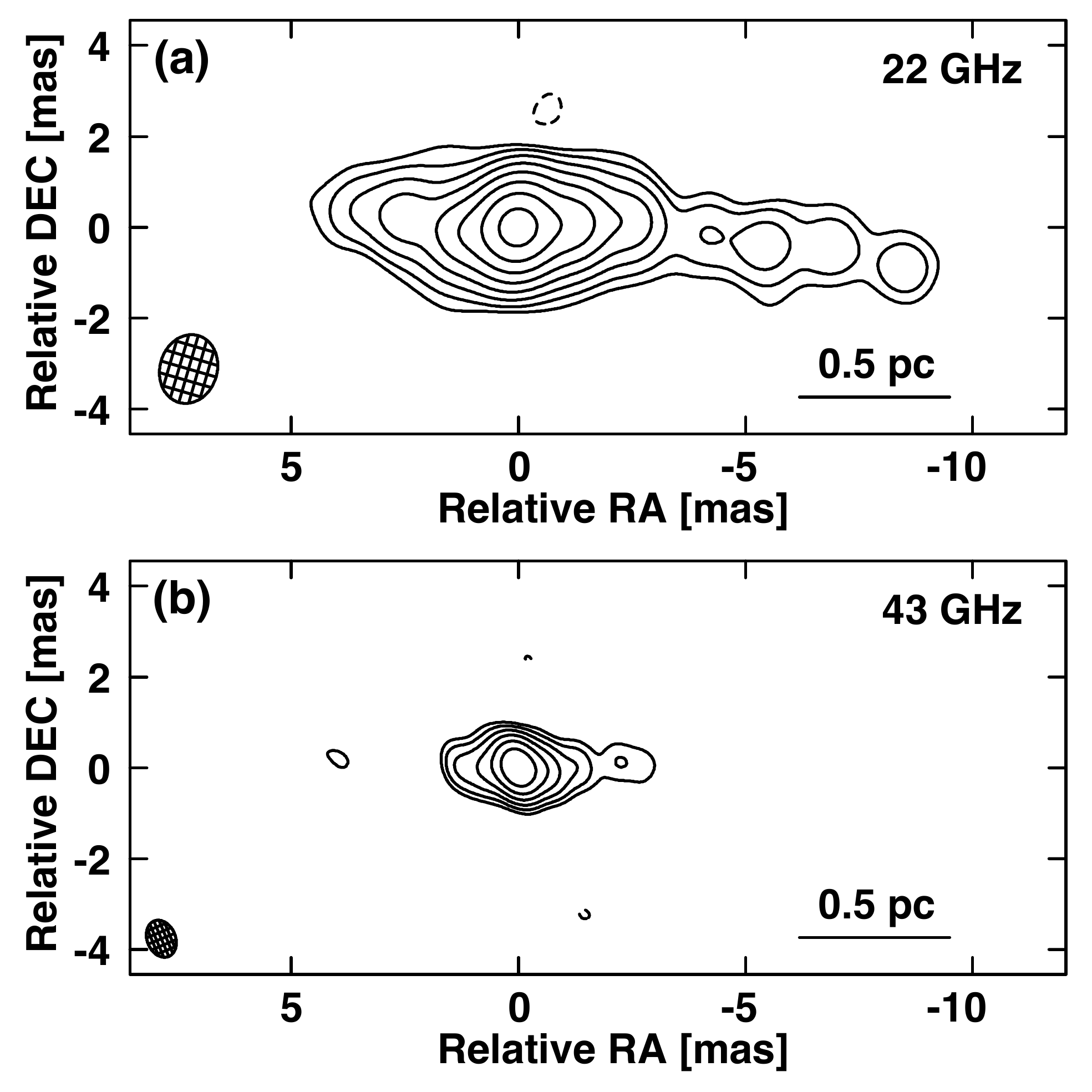} 
 \end{center}
\caption{
Continuum images of NGC~4261 at 
(a) 22~GHz and (b) 43~GHz. 
The coordinate origin is set at 
the brightest position for each image. 
Contours begin at $\pm$ four times $I_\mathrm{rms}$ 
and increase by a factor of two.
The synthesized beam is indicated 
by a cross-hatched 
ellipse at the bottom-left corner 
of each image. 
}
\label{fig:cntmap}
\end{figure}

Figure~\ref{fig:cntmap} 
presents continuum images of the 
pc-scale jets 
of NGC~4261 at 22 and 43~GHz. 
The image parameters are summarized  
in table~\ref{tab:obsinfo}. 
At both frequencies,
the two-sided jet structure and
a central bright core (base of the jet) 
appear to be aligned 
along the direction of the position angle (PA) 
$86\pm2^{\circ}$, 
which is in agreement 
with previous estimations of the jet axis 
ranging from kpc to pc scales 
\citep{birkinshaw85, jones00, piner01, sss22}. 

\begin{figure}
 \begin{center}
  \includegraphics[width=80mm]{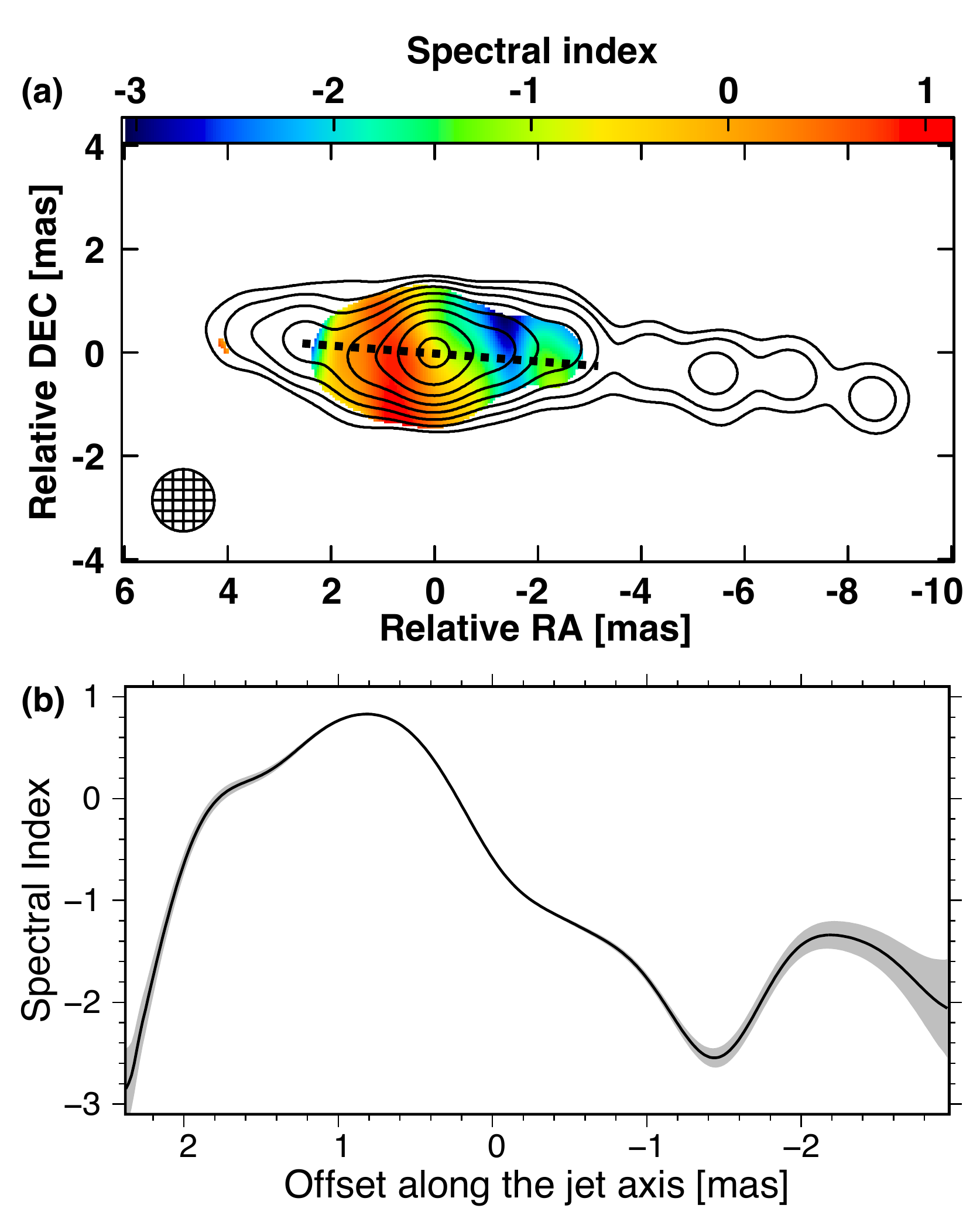} 
 \end{center}
\caption{
(a) 
Color-scale spectral index distribution 
between 22 and 43~GHz, 
superimposed on the 22~GHz continuum image 
restored with a circular Gaussian beam of 1.2 mas. 
The coordinate origin is set to 
the brightest position of the restored 22~GHz continuum image. 
Contours begin at four times $I_\mathrm{rms}$ and  
increase by a factor of two.
The synthesized beam is indicated 
by a cross-hatched 
ellipse at the bottom-left corner 
of each image. 
The dashed black line indicates 
the jet axis direction (PA = 86$^{\circ}$). 
(b) 
Slice profile of the spectral index 
along the jet axis 
(black dotted line in (a)). 
Gray-shaded area represents the 1-$\sigma$ error region.  
}
\label{fig:spixmap}
\end{figure}

At 22~GHz, 
a bright nuclear source and a faint jet-like emission 
were detected extending over 14 mas (2.1~pc) 
along the jet axis. 
The bright nuclear source was partially resolved 
into a core component and 
several nuclear jet components on the both sides  
spanning 8 mas (1.2~pc) from east to west. 
In addition to the nuclear source, 
the jet-like emission extended 9 mas (1.4~pc) 
along the west from the brightest position,
and appeared to consist of 
several weak jet components. 

In the 43~GHz map, 
the core was more prominent, and 
the jet components appeared to be weaker or undetectable. 
The continuum source showed a bright core component 
with sub structures on the both sides, 
spanning 5 mas (0.76~pc). 
The jet component observed on the western approaching side 
was clearly resolved from the core component. 
In addition, a faint jet component was slightly visible  
on the eastern receding jet side, 
4 mas east of the brightest position.

\subsection{Spectral Index}

We analyzed the distribution of 
the spectral index $\alpha$ 
($S_{\nu} \propto \nu^{\alpha}$)
between 22 and 43~GHz.  
For this, 
we first restored both continuum images 
with a 1.2 mas circular Gaussian beam, 
which denotes the minor-axis size of the synthesized beam 
at 22~GHz.  
Following this, we determined 
the position of the optically thin nuclear jet component,  
as the reference position for each map, 
based on Gaussian model fitting. 
The reference nuclear jet component 
was located 2.5 and 2.8 mas east 
of the brightest components  
at 22 and 43~GHz, respectively. 
We aligned the restored images, 
based on the reference positions. 

The spectral index distribution derived 
from the restored images, 
and a slice in the spectral index along the jet axis 
are depicted 
in figure~\ref{fig:spixmap}(a)
and (b), respectively. 
An inverted spectrum ($\alpha > 0$) is observed   
just east of the brightest position at 22~GHz. 
The maximum value of the spectral index is $\alpha = 0.83$
at the position 0.8 mas east of the brightest position 
at 22~GHz. 
The region with $\alpha > 0$ spans 1.8 mas along the jet axis. 
For the western nuclear jets, 
the spectral index decreases and 
drops down to $\alpha < -2$ at 1.5 mas west 
of the brightest position. 
The spectral index increases again, and yet 
another peak of the spectral index ($\alpha = -1.3$) 
can be seen at the western nuclear component, 
2.2 mas west of the brightest position. 
This variation in the spectral index along the jet axis
including the second peak in the western nuclear jet is,  
agrees with the results of 
the Very Large Baseline Array  
at 22 and 43~GHz 
\citep{haga15}.

\begin{figure*}
 \begin{center}
  \includegraphics[width=0.95\linewidth]{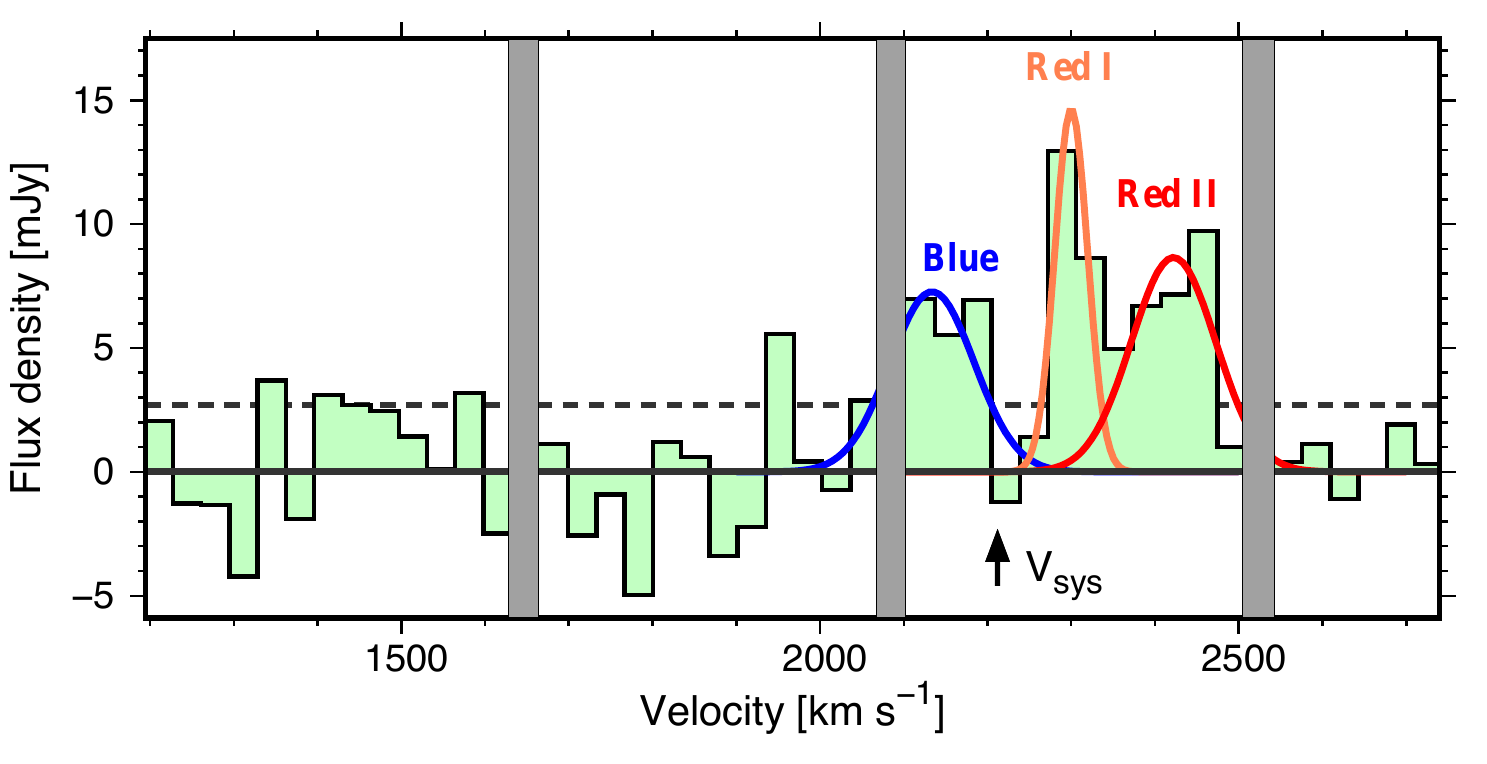} 
 \end{center}
\caption{
Extracted H$_2$O maser line spectra 
integrated over 
a region of 3$\times$3 mas 
centered 
at ($\Delta$RA, $\Delta$DEC)=(1.5 mas, 0.0 mas). 
The spectra are binned to 33.7 km s$^{-1}$ 
per channel, and the continuum emission 
is subtracted. 
The dashed black line indicates 
an rms noise density of 2.7 mJy 
in flux density. 
Gray rectangles indicate 
the flagged frequency channels 
at the the edges of each SW. 
The blue, orange and red curves represent   
a triple-Gaussian fitting of the spectral profile, 
and the fitting results are summarized 
in table~\ref{tab:gauss}. 
}
\label{fig:spc}
\end{figure*}

\begin{table*}
  \tbl{H$_2$O maser spectral parameters}{%
  \begin{tabular}{lcccccc}
    \hline
    Component 
    & $S_\mathrm{p}$\footnotemark[$\ddag$]
    & FWHM 
    & $V_\mathrm{ctr}$\footnotemark[$\S$] 
    & $V_\mathrm{ctr}-V_\mathrm{sys}$  
    & \multicolumn{2}{c}{$t$-test for $S_\mathrm{p}$}  \\
 \cline{6-7}   
    &   &   &   &   & $t$-value  & $p$-value \\ 
    & (mJy) & (km~s$^{-1}$) & (km~s$^{-1}$) & (km~s$^{-1}$) &  & \\
      \hline
\multicolumn{7}{c}{Triple-Gaussian Fit} \\
      \hline
    Blue & $7.3\pm2.0$ & $118\pm38$ & $2134\pm19$ & $-78$  & 3.71 & 0.0007 \\
    Red $\mathrm{I}$ & $14.6\pm4.5$ & $46\pm19$  & $2300\pm6$ & $+88$ & 3.22 & 0.0028 \\
    Red $\mathrm{I}\hspace{-1.0pt}\mathrm{I}$ & 
  $ 8.7 \pm 2.0$ & $119 \pm 39$  & $2422 \pm 14$ & $+210$ & 4.39 & 0.0001 \\
    \hline
\multicolumn{7}{c}{Double-Gaussian Fit} \\
    \hline
    Blue & $7.0\pm2.5$ & $105\pm45$ & $2124\pm23$ & $-89$  & 2.76 & 0.0089 \\
    Red & $8.5\pm1.6$ & $221\pm63$  & $2365\pm23$ & $+153$ & 5.18 & 8e-06 \\
    \hline
  \end{tabular}}
\label{tab:gauss}
\begin{tabnote} 
\footnotemark[$\ddag$]  Peak flux density.  \\
\footnotemark[$\S$]  Centroid velocity.  \\ 
\end{tabnote}
\end{table*}


\subsection{H$_2$O masers}

The spectral profile of H$_2$O maser emission 
from the pc-scale region in NGC~4261, 
obtained from the VLBI data, 
is presented in figure~\ref{fig:spc}. 
H$_2$O maser features are 
detected above the 3 $\sigma$ level 
at the channels of 2289--2323 
and 2458 km~s$^{-1}$ in velocity. 
The peak flux density is 12.9 mJy at 2289 km~s$^{-1}$. 
In addition to these, 
a possible feature detection 
at the 2--2.5 $\sigma$ level can be observed 
at the velocity channels of 2121--2188 km~s$^{-1}$ 
and 2357--2424 km~s$^{-1}$. 
The spectral profile appears to be resolved into 
two different velocity groups, 
and it resembles the H~$\textsc{i}$ profile 
with two peaks distributed around 2170 and 2250 km~s$^{-1}$,  
as revealed by EVN observations  \citep{vanlangevelde00}. 
The red-shifted velocity group exhibits 
a double-peaked profile 
at 2289 and 2458 km~s$^{-1}$, 
and this group could consist of two velocity components.
However, the previous GBT observations have shown 
a broad single feature \citep{wagner13}, 
and 
the observed differences in the spectral profile 
could be attributed to time variation 
or 
resolving out emissions on the VLBI baselines. 
Further, we applied a triple-Gaussian fit  
to the spectrum, depicted in figure~\ref{fig:spc}. 
For this Gaussian fit, 
the reduced $\chi^2$ value of 2.486 for 34 degrees of freedom. 
However, 
a tentative double-Gaussian fit to the spectrum results in 
a reduced $\chi^2$ value of 2.854 on 37 degrees of freedom, 
which is slightly worse than that for a triple-Gaussian fit. 
The fitting results are summarized 
in table~\ref{tab:gauss}. 
To further examine the significance of the maser line detection, 
we performed a $t$-test 
by setting a null hypothesis of $S_\mathrm{p}$ = 0. 
The $t$-test yielded a significance of $S_\mathrm{p}$ for each velocity component. 
The isotropic luminosity ($L_\mathrm{H2O}$) 
was estimated to be 60 $L_{\odot}$, 
based on the following formula:  
\begin{equation}
L_\mathrm{H2O} = 1.04 \times 10^{-3} 
\nu_\mathrm{rest} D_\mathrm{L}^2 \int S dv
 ~ L_{\odot} , 
\end{equation}
where $\nu_\mathrm{rest}$ is 22.235 GHz, 
and 
$\int S dv$ 
is assumed to represent the sum of 
$S_\mathrm{p} \times$ FWHM 
in Jy~km~s$^{-1}$ for each component 
listed in table~\ref{tab:gauss}. 
The derived $L_\mathrm{H2O}$ is close to 
the value estimated by GBT observations \citep{wagner13}.

\begin{figure*}
\begin{center}
\includegraphics[width=0.9\linewidth]{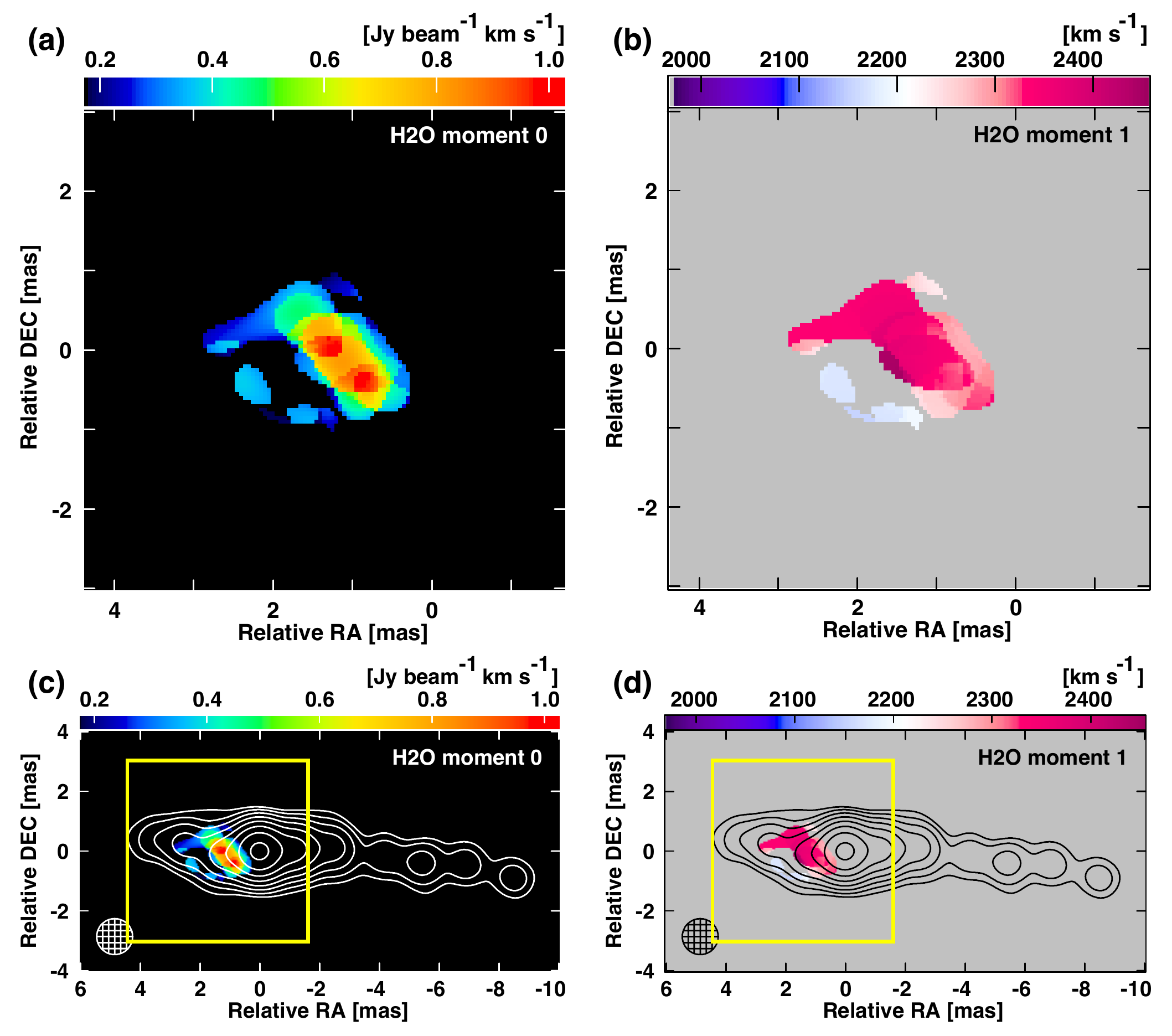} 
\end{center}
\caption{
Close-up view of the 
(a) velocity-integrated intensity (moment 0) 
and (b) intensity-weighted velocity (moment 1) maps  
of the H$_2$O maser. 
Relative distributions 
of (c) the moment-0 and (d) moment-1 maps 
with respect to the 22~GHz continuum image 
restored with a circular Gaussian beam of 1.2 mas 
(contours). 
The yellow squares in (c) and (d) represent  
the areas of (a) and (b), respectively. 
The color range in the moment-1 maps denotes  
$\pm240$ km~s$^{-1}$ 
from $V_\mathrm{sys}$. 
}
\label{fig:momnt}
\end{figure*}

Further, we generated maps of 
velocity-integrated intensities (moment 0) and 
intensity-weighted velocities (moment 1) 
with respect to the 22~GHz continuum source 
(figure~\ref{fig:momnt}), 
by clipping out the pixels 
where the H$_2$O maser emission 
was lower than 1.5 $\sigma$ in the image cube. 
The moment-0 maps displayed in figure~\ref{fig:momnt}(a)(c) 
reveal a prominent elongated structure 
 along a PA of 50$^{\circ}$ 
just east of the brightest continuum position at 22~GHz, 
i.e., upstream in projection 
from the base of the western approaching jet. 
We determined the position 
of the prominent structure 
in the moment-0 map 
using a two-dimensional Gaussian fit. 
Consequently, the derived relative position   
from the brightest continuum position 
was found to be 
($\Delta$RA, $\Delta$DEC) 
$=$ ($+$1.12 mas, $-$0.08 mas).

The moment-1 maps displayed in figure~\ref{fig:momnt}(b)(d)
indicate that the red-shifted emission 
originates from the elongated structure.
In addition, a few faint clumps can be observed 
southeast of the elongated structures. 
The possibly detected blue-shifted emission 
could be associated with the faint clumps,  
while the clumps do not emerge at the 3 $\sigma$ level 
in their channel maps.


\section{Discussions}

\begin{figure*}
 \begin{center}
  \includegraphics[width=0.90\linewidth]{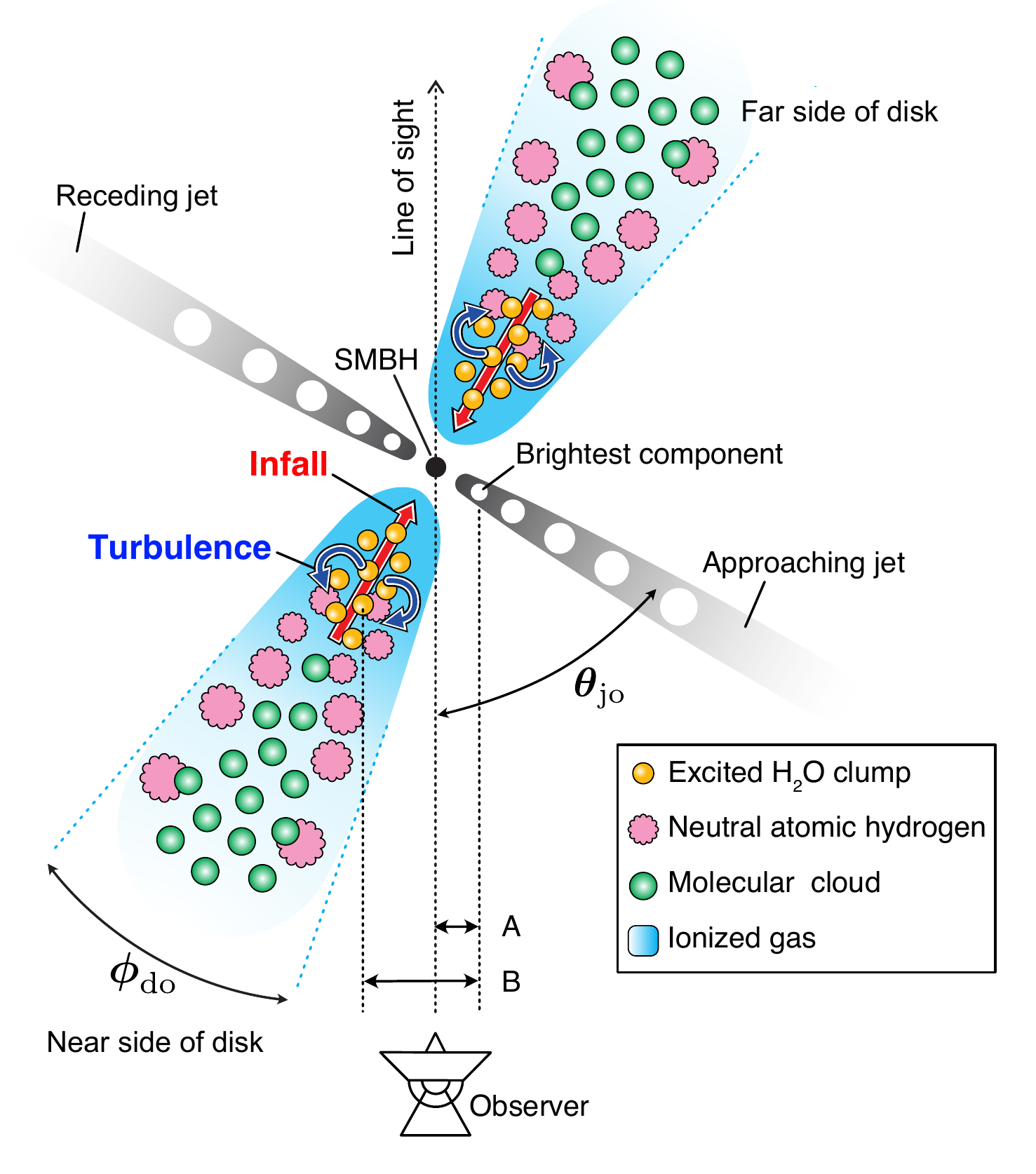} 
 \end{center}
\caption{
Possible geometry 
of the inner pc-scale region in NGC~4261, 
showing the two-sided jet, the SMBH and 
the obscuring disk. 
The jet axis is inclined 
at an angle of $\theta_\mathrm{jo}$ 
with respect to the line of sight. 
The obscuring disk consists of the 
ionized and molecular gases, 
with a full opening angle of $\phi_\mathrm{do}$. 
A and B denote the angular separations  
of the SMBH and the H$_2$O maser 
relative to the brightest continuum component, 
respectively. 
The ionized gas region is formed 
on the inner surface of the disk. 
Excited H$_2$O maser clumps are located 
next to the ionized gas region, 
at sub-pc radii. 
Cooler molecular clouds 
are distributed at outer radii up to 100 pc. 
H~$\textsc{i}$ gas resides at pc radii, 
and may extend to outer radii. 
A complex combination of several motions 
such as infall, turbulence and outflow 
produces a broad line profile.  
Past VLBI observations have indicated that 
$\theta_\mathrm{oj}$ 
and $\phi_\mathrm{do}$ are  
($63\pm3$)$^{\circ}$ and 
$\sim13^{\circ}$, 
respectively. 
}
\label{fig:model}
\end{figure*}


\subsection{Disk Scenario}

In this section, we discuss the association of the H$_2$O maser 
with the FFA obscuring disk in NGC~4261. 
As stated, H$_2$O maser emission could be detected 
1.12 mas east of the brightest continuum position 
at 22~GHz, 
where the spectrum is inverted ($\alpha >$ 0).
This indicates that 
the H$_2$O maser clumps spatially coincide 
with the ionized gas 
because a highly inverted spectrum ($\alpha >$ 2.5) 
has already been observed in the same region 
at the lower frequencies 
and interpreted 
as the FFA by the foreground ionized gas in a part of 
the obscuring disk 
\citep{jones97, jones00, jones01, haga15}. 
A possible schematic view of the FFA disk 
is shown in figure~\ref{fig:model}. 
The inner surface of the disk is directly illuminated 
by X-ray radiation 
originating from the central source, 
and an extended layer of hot ionized gas is generated 
on the surface. 
The SMBH and the base of the receding jet are 
obscured by the foreground ionized gas. 

Molecules can survive inside the disk 
at lower temperatures, 
and form two regions 
under different physical conditions. 
Excited H$_2$O molecules with an inverted population 
lie in the XDR, 
partially illuminated by the X-ray radiation,
close to the ionized surface. 
On the other hand, 
cooler molecules 
without an inverted population 
lie in the outer region  
up to hundreds pc, 
probing a rotation disk 
as revealed by the interferometric observations 
\citep{boizelle21, sss22}. 
The XDR should be formed on both the near and 
far sides of the disk. 
The absence of spike-like narrow maser lines suggests 
the low-gain maser amplification 
under saturated conditions. 
Thus, luminous H$_2$O maser emissions require 
the continuum seed emission. 
Excited H$_2$O molecules 
in the XDR of the near side 
amplify the background continuum seed emission 
from the receding jet, 
and produce the luminous H$_2$O maser emission. 
By contrast, 
the continuum seed emission is not adequate 
for the XDR on the far side 
behind the approaching jet,  
resulting in extremely weak or 
undetectable H$_2$O maser emission 
on the approaching jet side. 
If the position of the SMBH is 0.20 mas east 
of the brightest continuum 
at 22~GHz in projection, or A = 0.20 mas in figure~\ref{fig:model}, 
the the angular separation of the H$_2$O maser 
from the SMBH is B = 0.92 mas, 
projected on the eastern receding radio jet. 
Assuming that the disk rotation axis, that is, the jet axis,  
is inclined at 64$^{\circ}$ 
relative to the line of sight \citep{jaffe96},  
the location of 
the H$_2$O maser emission is determined to be 
at a radius of 0.32 pc in the disk. 

The velocities of H$_2$O maser emission 
($V_\mathrm{sys} \pm 200$ km~s$^{-1}$)  
at sub-pc radii 
are inconsistent with 
the tangential velocity of the 100-pc CND rotation,  
as revealed by the thermal molecular lines.  
This is because the tangential velocity 
at a radius of 0.32 pc 
should ideally be $10^4$ km~s$^{-1}$, 
given the enclosed mass of 
1.6$\times$10$^{9}$ $M_{\odot}$ 
\citep{boizelle21, sss22}.  
The broad H$_2$O maser spectrum 
could trace 
various kinematics 
lying at sub-pc radii, 
close to the apparent SMBH location. 
The red-shifted H$_2$O maser feature  
is significant, and
the ongoing infall motion toward the SMBH 
can simply explain 
the red-shifted velocity. 
The blue-shifted maser feature could be 
attributed to 
outward motion from the disk, 
such as outflow and/or turbulence 
driven by AGN radiation fields 
(e.g., \cite{wada12}). 

Previously, H~$\textsc{i}$ gas has been 
interpreted to originate from a disk 
\citep{vanlangevelde00}. 
As the angular separation between the 
core component at 1.4 GHz and the SMBH location 
was measured to be 5.0 mas ($4.92+0.082$ mas) 
by \citet{haga15},
the mean radius of the H~$\textsc{i}$ structure 
is calculated to be 30 mas or 4.5 pc,   
given a $\theta_\mathrm{jo}$ value of 64$^{\circ}$.
The similarity between the H$_2$O maser 
and H~$\textsc{i}$ spectra 
indicates that 
the H$_2$O maser and H~$\textsc{i}$ 
could probe the same complex kinematics 
for the sub-pc and pc radii 
in the disk. 

Thus, all the observed characteristics 
can be naturally 
explained by the multi-phase disk scenario. 
A spatial coincidence between an ionized gas 
and the red-shifted H$_2$O maser gas 
has also been found in NGC~1052, 
which hosts a multi-phase gas torus 
with several layers of 
hot innermost ionized gases, 
warmed H$_2$O molecular clouds,
and cooler molecular clouds such as HCN and HCO$^+$
\citep{sss08, sss16, sss19}. 
Because a narrow maser line feature has been found 
in NGC~1052 \citep{kameno05}, 
the NGC~4261 could exhibit lower-gain maser emission 
compared with NGC~1052.


\subsection{Jet Interaction Scenario}

An alternative explanation 
for the origin of H$_2$O maser emission 
is the jet interaction 
with ambient H$_2$O molecular clouds. 
In this scenario, 
the H$_2$O maser emission can be expected 
to originate from the shocked region 
surrounding the eastern receding jet  
by amplifying the continuum seed emission 
from the jet. 
This scenario explains  
the association of red-shifted H$_2$O maser 
with the receding jet. 
The blue-shifted H$_2$O maser spectrum 
could be attributed to  outward kinematics 
such as turbulence or outflow. 
The jet interaction scenario is consistent with 
the absence of narrow maser lines. 
The propagation speed 
of the shock front $v_\mathrm{s}$ is defined as
\begin{equation}
    v_\mathrm{s} = v_\mathrm{j} 
    \Bigl(
    \frac{\rho_\mathrm{0}}{\rho_\mathrm{j}}
    \Bigr)^{-1/2},
\end{equation}
where $v_\mathrm{j}$ denotes the jet velocity  
and 
$\rho_\mathrm{0}/\rho_\mathrm{j}$ denotes  
the density ratio between the jet $\rho_\mathrm{j}$ 
and the ambient molecular gas 
$\rho_\mathrm{0}$
(e.g., \cite{kameno05}). 
Assuming $v_\mathrm{j}$ = 0.46 $c$ and 
$v_\mathrm{s}$ = 149 km~s$^{-1}$ 
(the mean velocity of Red $\mathrm{I}$ and 
Red $\mathrm{I}\hspace{-1.0pt}\mathrm{I}$), 
the density ratio can be calculated to be 
$\rho_\mathrm{0}/\rho_\mathrm{j} = 8.6\times10^5$. 
If we apply an H$_2$ number density of 10$^7$--10$^9$ 
cm$^{-3}$ 
to $\rho_\mathrm{0}$, 
based on the requirements for population inversion 
of the 22~GHz H$_2$O maser transition 
in a shock 
\citep{kaufman96,hollenbach13},
$\rho_\mathrm{j}$ would range from 
10$^{-23}$ to 10$^{-21}$ g~cm$^{-3}$.
The jet kinetic power $P_\mathrm{j}$ is defined as 
\begin{equation}
P_\mathrm{j} = 
\frac{\pi}{2}\rho_\mathrm{j} v_\mathrm{j}^3 r_\mathrm{j}^2
\end{equation}
where $r_\mathrm{j}$ denotes the jet radius. 
Adopting $r_\mathrm{j}$ = 0.05 pc 
($\sim 10^3$ Schwarzschild radius)
based on the sub-pc jet width measurement 
by \citet{nakahara18}, and 
$v_\mathrm{j}$ = 0.46 $c$ 
and $\rho_\mathrm{j}$ = 10$^{-23}$--10$^{-21}$ g~cm$^{-3}$
from the above discussion, 
the resultant $P_\mathrm{j}$ is 
10$^{42}$--10$^{44}$ erg~s$^{-1}$. 
This value is in agreement with 
the jet power obtained from 
the spectral energy distributions modeling 
conducted by \citet{demenezes20}. 

However, the jet interaction scenario 
leaves unexplained 
the explicit association of the H$_2$O maser emission 
with the receding jet 
where the continuum spectrum is optically thick, 
despite the maser association with the 
optically thin part of the approaching jet 
in the proposed jet-excited megamasers 
such as 
Mrk 384 \citep{peck03} and 
IRAS 15480-0344 \citep{castangia19}. 
Thus, further high-sensitivity VLBI imaging 
would be extremely helpful in 
accurately revealing 
the location of the H$_2$O maser clumps 
in the jet structure 
and 
in better elucidating 
the complex gas kinematics 
in the immediate vicinity of the SMBH.



\section{Conclusions}

We conducted high-sensitivity VLBI observations 
of the second-nearest radio-loud H$_2$O megamaser source
NGC~4261 at 22~GHz and 43~GHz. 
The captured double-frequency VLBI images 
of the continuum emission 
have revealed a 
two-sided jet structure and a central bright component
aligned along the east--west direction. 
The resultant spectral index distribution 
along the jet structure was found to be 
consistent with the FFA on the jet bases 
caused by the dense ionized gas associated with the obscuring disk. 
H$_2$O maser emission was detected 
at a slightly red-shifted velocity 
relative to $V_\mathrm{sys}$ 
with a significance 
over the 3 $\sigma$ level,  
utilizing the in-beam phase-referencing technique.
Further, we imaged the H$_2$O maser emission, 
at positions where the FFA opacity was high 
on the eastern receding jet. 
Our results suggested that 
the H$_2$O maser emission in NGC~4261 
could be 
associated with the inner radius of 
the obscuring disk, 
as it was proposed 
for the nearest radio-loud megamaser source 
NGC~1052. 
An alternative scenario 
on the H$_2$O maser association is 
the shock region of the interaction 
between the jet and ambient molecular clouds. 
However, this does not explain 
why the H$_2$O maser emission is associated with 
the receding jet only, 
at positions 
where the continuum spectrum is optically thick.



\begin{ack}
We acknowledge all members 
at KVN, VERA and Ibaraki University 
who supported the operation of the array 
and the correlation of the data. 
KVN is a facility operated 
by the Korea Astronomy and Space Science Institute (KASI), 
and VERA is a facility operated 
by the National Astronomical Observatory of Japan (NAOJ) 
in collaboration with associated universities in Japan. 
The Takahagi 32-m telescope is operated 
by NAOJ and Ibaraki University and 
partially supported by the Inter-university 
collaborative project “Japanese VLBI Network (JVN)” of NAOJ. 
S.S.-S. is supported by JSPS KAKENHI grant No. 21K03628.
N.K. is supported by JSPS KAKENHI grant No. 19K0391.
\end{ack}





\end{document}